# Precessional switching of thin nanomagnets: analytical study

T. Devolder and C. Chappert.

Institut d'Electronique Fondamentale, UMR CNRS 8622, Bât. 220, université Paris-Sud, 91405 ORSAY, France.

*Abstract : We study analytically the precessional switching of the magnetization of a thin macrospin. We analyze its response when subjected to an external field along its in-plane hard axis. We derive the exact trajectories of the magnetization. The switching versus non switching behavior is delimited by a bifurcation trajectory, for applied fields equal to half of the effective anisotropy field. A magnetization going through this bifurcation trajectory passes exactly along the hard axis and exhibits a vanishing characteristic frequency at that unstable point, which makes the trajectory noise sensitive. Attempting to approach the related minimal cost in applied field makes the magnetization final state unpredictable. We add finite damping in the model as a perturbative, energy dissipation factor. For a large applied field, the system switches several times back and forth. Several trajectories can be gone through before the system has dissipated enough energy to converge to one attracting equilibrium state. For some moderate fields, the system switches only once by a relaxation dominated precessional switching. We show that the associated switching field increases linearly with the damping parameter. The slope scales with the square root of the effective anisotropy. Our simple concluding expressions are useful to assess the potential application of precessional switching in magnetic random access memories.*

Obtaining reproducible magnetization switching within the sub-nanosecond regime and the sub-micron range is currently one of the most challenged tasks in nanomagnetism [1]-[6]. An intense research activity is currently in progress for measuring [5], [7], [8], [9] the ultrafast dynamics of nanomagnets and for accounting for it numerically [10]. In Magnetic Random Access Memories, it is of paramount industrial interest to design magnetization switching strategies, that succeed to commute the storage elements magnetization in a fast, reliable and energy cost-effective way that can be easily scaled down.

The conventional strategy to switch magnetization is to apply an external field **H** antiparallel to the magnetization. This strategy is cost effective in quasi-static reversal, when thermal activation helps to overcome the energy barriers [11] i.e. at times longer than a few nanoseconds. In the faster regime, an external field applied antiparallel to **M** creates no torque on the magnetization, except at the few places where the magnetization is not strictly along the easy axis. The first consequence is that the magnetization rotation starts slowly. The resulting reversal time is typically 2-5 ns [7], and reducing this time is very expensive in applied field [12]. The second (and worst) consequence is that this strategy is cannot be scaled down in size. It will be even less effective for smaller elements because their magnetization will tend to more uniformity, which reduces the area undergoing a finite torque, and consequently the switching speed.

In 1996, the seminal work of He et al. [14] studied the effect of fast-rising fields onto the magnetization of macrospins (i.e. magnetic bodies with perfectly uniform magnetization). They predicted using numerical computations that fields non antiparallel to the magnetization, thus triggering precessional motions of the magnetization vector, could induce fast switching events for fields smaller than the Stoner-Wohlfarth criterion [14]. This was confirmed in more details by several authors [13], [15]. The lowest switching field was predicted to be half of the anisotropy field ($H_K$). The effect of finite rise-time and finite damping were studied [6], [13] by numerical integration of the Landau-Lifschitz-Gilbert (LLG) equation [16]. Deep sub-ns precessional switching events, lasting half a precession period (typically 200 ps) were experimentally confirmed last year [2]-[4], [9]. Detailed experimental studies [4] indicated that precessional switching can be achieved at sub-Stoner-Wohlfarth fields on "large" (2×5 μm², i.e. non macrospin) particles, but with a slightly less favorable minimal applied field. The reproducibility was questioned when attempting to reach the promised minimal field cost $H_K/2$ [4]. The direct-write reliability issue was studied for larger samples in the cross-wire configuration mimicking a MRAM architecture in [9].

In this paper, we report an analytical analysis of the precessional switching of soft nanomagnets. Some very enlightening information has already arisen from numerical integration of LLG equations [15], [6]. However, because of multiple parameters intricacy (damping, field magnitude and orientation, anisotropy, saturation magnetization), only snapshots in the parameter space have been reported so far, obtained through numerical computations. The scope of this paper is to provide a fully analytical analysis of the precessional switching so as to derive the general behavior versus the full set of parameters. In doing so, we find the criteria for the success of a precessional switching strategy.

Section I is devoted to the magnetization trajectory when damping is neglected. We show how the magnetization trajectory deforms when the anisotropy is varied above $H_K/2$. The time evolution of the magnetization is slowed down by the anisotropy. It is found that the magnetization switches if a bifurcation criterion is satisfied, which depends on the initial magnetization. The bifurcation trajectory passes exactly through the hard axis which is a stationary and unstable magnetization position. It makes this trajectory very sensitive to any fluctuation, and fundamentally unpredictable when attempting to approach the minimal cost in applied field in realistic conditions.

In section II, the effect of damping is analyzed in a perturbative manner, and criteria are derived for the analytical assessment of the switching ability of a given set of experimental parameters. A special focus is dedicated to the relaxation dominated precessional switching, where the minimal cost in applied field is shown to increase linearly with the damping constant, the slope of this dependence scaling with the square root of the anisotropy. Finally, we make a thorough comparison with results obtained from direct time integration of the equation of motion.

Magnetization dynamics of a macrospin can be accurately described by the well-known Landau-Lifshitz equation [16]:

$$\text{Eq. 1} \qquad \frac{d\vec{M}}{dt} = \gamma_0 \vec{H}_{eff} \times \vec{M} - \frac{\alpha}{\|\vec{M}\|}\left[\frac{d\vec{M}}{dt} \times \vec{M}\right]$$

where $\gamma_0 = \gamma\mu_0$ and $\gamma/2\pi = 28$ GHz/T is the gyromagnetic factor [17], and the instantaneous effective field is the sum of the applied field **H**, the anisotropy field **H**$_K$ and the demagnetizing field **H**$_D$. For a macrospin, the magnetization **M** is uniform and the exchange field is zero. The relaxation towards equilibrium is described phenomenologically by the damping constant α. Throughout this paper, international SI units are used: **H** and **M** are in A/m, $\mu_0$**H** and $\mu_0$**M** are in Tesla, and $\gamma_0 H$ and $\gamma_0 M_S$ are frequencies.

The applied field points strictly along the *+(y)* direction (see inset in Figure 1A). We define $h = H/M_S > 0$ its reduced strength. We define **m**=**M**/$M_S$ the reduced magnetization, and $m_x$, $m_y$, $m_z$ its projections. The system has an easy axis along *(x)*, i.e. in the sample plane, assumed to arise from both shape and magneto-crystalline uniaxial anisotropy. The initial magnetization is assumed at rest exactly along the *+(x)* axis, except for Eq. 10 for which **m**$_{(t=0)}$ is arbitrary.

The magneto-crystalline anisotropy field is along *(x)* and it is $H_K = h_k M_S m_x$.

The system is an ellipsoid with principal axes along *(x)*, *(y)* and *(z)*, and corresponding demagnetizing factor $N_x$, $N_y$ and $N_z$. We are concerned about thin film macrospin, such that $N_x \leq N_y \ll N_z \approx 1$. We consider that $h_k + (N_y - N_x) > 0$, consistent with an easy axis along *(x)*.

For simplicity we write the equations with only two parameters that are $h_k$ and $N_z$. The demagnetizing field is written solely along (z) and is $H_{demag} = -N_z M_S m_z$. To take into account the in-plane components of the demagnetizing field, our equations can be straightforwardly generalized by systematically replacing $N_z$ by $N_z - N_y$ and $h_k$ by $h_k + (N_y - N_x)$. Exact expressions where we assume $h \ll N_z$ or $h_k \ll N_z$ are displayed using the symbol "≅". Approximate expressions implying other arguments use "≈".

## I. Magnetization trajectory with uniaxial in-plane anisotropy

The in-plane effective uniaxial anisotropy field is along the easy axis *(x)* and is $H_K = h_k M_S m_x$, such that the anisotropy energy is $-\frac{1}{2}\mu_0 M_S^2 \cdot h_k m_x^2$. The LLG equation with such anisotropy becomes:

$$\text{Eq. 2} \qquad \dot{m}_x = \gamma_0 M_S m_z \left( N_z m_y + h \right)$$

$$\text{Eq. 3} \qquad \dot{m}_y = -\gamma_0 M_S \left( N_z + h_k \right) m_z m_x$$

$$\text{Eq. 4} \qquad \dot{m}_z = -\gamma_0 M_S m_x \left( h - h_k m_y \right)$$

Substituting these equations together yields the magnetization trajectory. The latter can also be derived using the conservation of the magnetization norm and of the total energy at $t>0$:

$$\text{Eq. 5} \qquad \frac{E_{(t)}}{\mu_0 M_S^2} = \frac{1}{2}\left( N_z m_z^2 - h_k m_x^2 \right) - h m_y$$

Either of these two approaches yields the exact magnetization trajectory:

$$\text{Eq. 6} \qquad m_x^2 = 1 - \frac{2h}{N_z + h_k} m_y - \frac{N_z}{N_z + h_k} m_y^2$$

$$\text{Eq. 7} \qquad m_z^2 = \frac{2h}{N_z + h_k} m_y - \frac{h_k}{N_z + h_k} m_y^2$$

I.A. <u>Classification of the non damped trajectories</u>

Depending on the relative magnitudes of $h$ and $h_k$, three possible types of trajectory may occur. They are sketched in Figure 1. They depend on the position of the initial energy in the energy landscape. The simplest case (i) happens when the energy landscape has an absolute minimum at $m_y=1$.

(i) This is the case of rather high applied fields, i.e. when $h \geq h_k$. A typical trajectory is displayed in Figure 1 for $h = h_k = 0.01$. In this range of applied fields, the $m_z$ versus $m_x$ magnetization trajectory is a stadium-shaped trajectory. The maximum of $m_z$ is $m_z \cong \pm\sqrt{2h - h_k}$, and it is reached when $m_x=0$. The trajectory reaches systematically the full reversal, i.e. $m_x = -1$. The $m_z$ versus $m_y$ component is an open path which looks like a half ellipse. For such applied fields, the equilibrium magnetization is along *(y)*, and the trajectory rotates around that sole axis (Figure 1B, C).

(ii) When the applied field is such that $h_k > h > h_k/2$, i.e. smaller than the anisotropy field but still at least half of it, the energy landscape has a saddle point at $m_y=1$, and the initial energy is above it. The $m_z$ versus $m_x$ trajectory takes a bone shape (Figure 1B). It bends back to the easy plane when approaching the direction of the applied field. The trajectory reaches systematically the full reversal ($m_x = -1$). For such applied fields, there is two degenerate equilibrium magnetization states, that have $m_y = h/h_k$ and $m_x>0$ or $m_x<0$. Qualitatively, the trajectory rotates alternatively around each of them. For instance the maximum of $m_z$ is obtained when $m_y = h/h_k$ and takes the value:

$$\text{Eq. 8} \qquad m_z^{max} = h\frac{1}{\sqrt{h_k(N_z + h_k)}} \cong h/\sqrt{N_z h_k}$$

$m_z$ has a local minimum when $m_x=0$. At this point:

$$\text{Eq. 9} \qquad m_y^m = \sqrt{N_z^2 + N_z h_k + h^2} - h \text{ and } m_z \cong \sqrt{2h - h_k}$$

The $m_z$ versus $m_y$ is an open path which looks like a truncated ellipse (Figure 1A). The trajectories for applied fields $h > h_k/2$ will be referred hereafter as field-dominated trajectories, in contrast to the anisotropy-dominated trajectories of the following paragraph.

(iii) The situation is topologically different when the field is too low to allow switching, i.e. when it is less than half of the anisotropy field ($h < h_k/2$). In that case the initial energy is below that of the saddle point, such that the magnetization vector can never pass in the $m_x<0$ half space. Magnetization precesses about the *sole nearest* equilibrium magnetization position $m_y = +h/h_k$. $m_y$ oscillates between $0$ and $2h/h_k$. $m_x$ oscillates between $1$ and $\cong 1-4h^2/h_k^2$ and the $m_z$ versus $m_x$ trajectory is ovoid-like with a maximum radius of curvature is at $m_x=1$. The $m_z$ versus $m_y$ has an ellipsoidal shape (Figure 1A).

An important situation in real devices is that of an initial magnetization not at rest along *(x)*, for instance because the magnetization is still ringing as a result of a previous switching. Let us suppose now that the new initial magnetization is not along (x) as in I.A but is ($m_{x0}<1$, $m_{y0}$ and $m_{z0}$). The corresponding trajectories are simply obtained by replacing the variables {$m_x^2$-1, $m_y$, $m_y^2$, $m_z^2$} by the variables {$m_x^2$- $m_{x0}^2$, $m_y$-$m_{y0}$, $m_y^2$-$m_{y0}^2$, $m_z^2$-$m_{z0}^2$} in Eq. 6 and Eq. 7. The criterion separating anisotropy dominated and field dominated trajectories is then:

$$\text{Eq. 10} \qquad h_{bif} = (1+m_{y0})\frac{h_k}{2} + \frac{m_{z0}^2}{1-m_{y0}}\frac{N_z + h_k}{2}$$

Let us discuss the case $m_{z0}=0$. The trajectories with initial magnetization not along the easy axis differ from the above described behavior only when $h_k > h/(1+m_{y0}) > h_k/2$ (see Figure 2).
If $m_{y0}$ is negative, the switching is eased and is achievable at a reduced applied field. Qualitatively, it is because the systems feels a Zeeman torque creating the demagnetizing field during a longer time interval than when $m_{y0}=0$.
If $m_{y0}$ is positive, the Zeeman torque is reduced, which may hinder the switching. The maximum reduction is attained when $m_{y0}=h/h_k$, i.e. when initial magnetization is at the in-field equilibrium position. In that case, setting the field has strictly no effect onto the magnetization (see Figure 2, in the limit of $m_{y0}=0.625$).

I.B.  <u>Bifurcation, stationary and unstable point</u>

The situation of Eq. 10 (basically: $h = h_{bif} = h_k/2$) deserves a particular comment because it is a *<u>bifurcation</u>* trajectory. If the applied field is slightly higher, the system energy of above that of the hard axis and the trajectory is field dominated, i.e. very extended and $m_x$ reaches both *+1* and *-1*. If the applied field is slightly lower, the initial energy is below that of hard axis magnetization and the trajectory is anisotropy dominated and stays in the $m_x > 0$ half space.

It sets the *minimal field required for a precessionnal switching event*. This important result was already obtained [10], [15] by numerical integration of the LLG equation. Here we obtained it analytically.

This corresponds to a total adiabatic transfer of the initial anisotropy energy in the Zeeman energy. Neither demagnetizing field, nor anisotropy field are present at this bifurcation point. Since the applied field is parallel to **M**, the torque is zero. Note that even if the demagnetizing tensor incorporates non vanishing $N_x$ and $N_y$, this result still holds. As predicted by Acremann et al. [15], this bifurcation point is thus also a *stationary* point where $\dot{m}_{x,y,z} = 0$. In the absence of damping, a system fulfilling Eq. 10 will go *ballistically* (without ringing) to that point and stay there forever. Numerical integration of the LLG equation (Figure 3A) for anisotropy conditions very near the bifurcation criterion confirms a drastic slow down of the trajectory when approaching $m_x=0$.

Note that the required field $H_{bif}=H_K/2$ is half of the anisotropy field, i.e. only half of the field needed to align the magnetization along the hard axis in a quasi-static evolution. Hence, this point is stationary but *unstable*: in case of a small perturbation, the resulting torque repels the magnetization away from this point… In real systems, the thermal agitation or any source of magnetic noise or dispersion will kick magnetization away from this bifurcation-stationary point, and the final state will evolve randomly to either $m_x>0$ or $m_x<0$. This is the reason why poor switching reproducibility is obtained when one experimentally attempts to reach the promised minimal field cost $H_K/2$ as in ref. [4].

I.C.   Characteristic switching frequency

Since the bifurcation trajectory has a singularity in frequency ($\omega \to 0$), there is no constant typical frequency near $h \approx h_{bif}$. In this section, we search for the scaling laws of the switching frequency, more specifically in the case $h_k/2 < h < h_k$. To this aim we introduce three qualitative times that govern the order of magnitude of the switching frequency. The reader should consider these times as conceptual guides that will be useful in section II.B for the calculation of the final magnetic state with finite damping. We define the "initial delay" $\tau_1$ during which $m_x$ does not react much, the "slow-down time" $\tau_2$ during which the system stays near the hard axis, and "maximum speed time" $\tau_3$ the typical time during which there exists a large demagnetizing field.

Their definition is illustrated in Figure 3A, which displays $m_x(t)$ as computed from an exact numerical integration of the LLG equation.

First, it takes a time delay $\tau_1$ for the system to set its maximum demagnetizing field. Indeed, the initial rate of change of $m_x$ and $m_y$ is zero (see Eq. 2 and Eq. 3), which can be seen as an effective delay in the response of the in-plane projection of the magnetization to the applied field. In experiments sensitive to only $m_x$ and $m_z$, this time delay during which the sole $m_z$ changes significantly is never accounted for [4]. We define this initial delay $\tau_1$ such that: $\tau_1 \dot{m}_z\big|_{t=0} = m_z^{max}$. Using Eq. 8, we get:

$$\text{Eq. 11} \qquad \tau_1 \cong \frac{1}{\gamma_0 M_S} \frac{1}{\sqrt{h_k(h_k + N_z)}} \quad \text{with } h_k > h > h_k/2$$

Or $\tau_1 \cong \dfrac{\sqrt{2h - h_k}}{\gamma_0 M_S h}$ if $h > h_k$.

As shown in Figure 3B, the initial delay during which the in-plane projection of the magnetization does not react much is typically $\tau_1 = 50$ ps for common soft alloys.

On the other hand, the reversal of the $m_x$ component occurs mainly due to the demagnetizing field and is thus dominated by the value of $m_z$ when $m_x \approx 0$. We define the "slow down time" $\tau_2$ as the time it takes for $m_x$ to pass from $+0.5$ to $-0.5$ as a result of the demagnetizing field, so that $\tau_2$ is such that $\tau_2 \dot{m}_x\big|_{m_x=0} \approx 1$. Using Eq. 4 and Eq. 9, we get:

$$\text{Eq. 12} \qquad \tau_2 \cong \frac{1}{\gamma_0 N_z M_S} \frac{1}{\sqrt{2h - h_k}} \quad \text{with } h > h_k/2$$

Using a similar approach, we can convert the demagnetizing field of Eq. 8 in a frequency, such that we can write $\tau_3$ the typical time spent by the system around $m_z^{max}$. During this "maximum speed time" $\tau_3$, there exists the largest demagnetizing field, driving the magnetization motion in a fast manner.

$$\text{Eq. 13} \qquad \tau_3 = \frac{\sqrt{N_z h_k}}{2\gamma_0 M_S h}$$

Note $\tau_1$, $\tau_2$ and $\tau_3$ are only qualitative conceptual guides: each of them may need to be multiplied by a numerical prefactor of the order of one. Their numerical prefactors, especially the one of $\tau_3$, will be reconsidered in section II.D, by doing a feed-back comparison with exact numerical calculations of the minimal switching field when damping is finite.

From Figure 3B, it is straightforward to see that $\tau_2 > \tau_1 > \tau_3$ holds almost always (for $0.51 h_k < h < h_k$). This justifies our naming convention for these times.

In addition, the slow-down time $\tau_2$ tends to infinity near the bifurcation criteria, such that we expect the characteristic frequency to be limited by $\tau_2$ and thus to scale with $\gamma_0 \sqrt{M_S (H - H_K / 2)}$. The latter switching frequency $\omega$ was calculated numerically. Plots of $\omega^2$ versus $-h_k$ or $h$ (not shown) are linear and can be very satisfactory fitted by:

$$\text{Eq. 14} \qquad \omega \approx 0.847 \gamma_0 \sqrt{M_S (H - H_K / 2)}$$

To summarize, the analytical expressions derived in this section to describe magnetization motion of a non damped, anisotropic macrospin thin film can account exactly for its magnetization trajectories. Those are either field dominated and allow switching when $h > h_k/2$, or either anisotropy dominated with no switching. At moderate switching fields (i.e. $h < h_k$), the magnetization rotates qualitatively in three steps $\tau_2 > \tau_1 > \tau_3$, which are respectively the slow-down time $\tau_2$, the initial delay $\tau_1$, and the maximum speed time $\tau_3$. The characteristic switching frequency tends to zero when $H = H_K/2$ and scales with the square root of the excess field above the $H_K/2$.

## II. Perturbation treatment of finite damping

We now evaluate the effect of finite damping. Since our aim is to access the technological potential of precessional switching, we restrict the analysis to field–dominated trajectories ($h > h_k/2$), i.e. to those trajectories where energy considerations do not forbid the switching event. Because we aim at defining cost-effective reversal strategies, we restrict to moderate fields ($h < h_k$).

Introducing damping into LLG equation makes it impossible to solve analytically. Therefore, we introduce damping a posteriori as the rate at which the energy in dissipated. With finite damping, the final magnetization at $t=+\infty$ will obey $m_y = h/h_k$ and $m_z=0$. In between initial ($m_x=+1$) and final state, the system bifurcates when the total energy gets smaller than that of the saddle point of the energy landscape. This arises at the time when an energy $\Delta E_{bif}$ has already been dissipated:

$$\text{Eq. 15} \qquad \frac{\Delta E_{bif}}{\mu_0 M_S^2} = h - h_k/2$$

The following paragraph (II.A) aims at evaluating the energy loss per cyclic trajectory $\Delta E_{cycle}$ and then to deduce the Number of Trajectories $NT = \Delta E_{bif}/\Delta E_{cycle}$ being gone through before the system bifurcates to an anisotropy-dominated trajectory and gets trapped in one of the $m_x>0$ or $m_x<0$ half spaces.

II.A. <u>Energy loss per unit cycling of the trajectory</u>

From [22] the damping is linked to the decay rate of the energy, i.e.:

$$\text{Eq. 16} \qquad \frac{1}{\mu_0 M_S^2} \frac{dE}{dt} = -\alpha \left( \dot{m}_x^2 + \dot{m}_y^2 + \dot{m}_z^2 \right) \frac{1}{\gamma_0 M_S}$$

When $m_x$ is 0 or $\pm 1$, the decay rates are found using Eq. 4, and respectively Eq. 2, Eq. 6 and Eq. 7:

$$\text{Eq. 17} \qquad \frac{1}{\mu_0 M_S^2} \frac{dE}{dt}\bigg|_{m_x=1} = -\alpha \gamma_0 M_S h^2 \text{ and } \frac{1}{\mu_0 M_S^2} \frac{dE}{dt}\bigg|_{m_x=0} = -\alpha \gamma_0 M_S N_z^2 (2h - h_k)$$

But most of the dissipation is likely to occur during the fastest parts of the trajectory, i.e. when the demagnetizing field is maximum during $\tau_3$. At that point, both $\dot{m}_x$ and $\dot{m}_y$ are both large, and $m_x^2 \cong 1 - h^2/2h_k^2$. Using $N_z \gg h$, Eq. 16 leads to:

$$\text{Eq. 18} \qquad \frac{1}{\mu_0 M_S^2} \frac{dE}{dt}\bigg|_{m_z^{max}} \cong -\alpha \gamma_0 M_S h^2 / h_k$$

We use $4\tau_1$ (Eq. 11), $2\tau_2$ (Eq. 12) and $4\tau_3$ (Eq. 8) for the typical time [23] intervals where the dissipations of Eq. 17 and Eq. 18 occur. Although the numerical factors will be justified a posteriori in section II.D, we mention that the 4 factor before $\tau_1$ corresponds to the initial delay, the delay when $m_x$ approaches $-1$, the following delay to set the opposite demagnetizing field, and the final slow down when $m_x$ approaches back to $+1$. The 2 factor before $\tau_2$ corresponds to back and forth reversal. The factor 4 in $4\tau_3$ correspond to the number of occurrences of $m_z^{max}$ in the case $h < h_k$. The total energy loss for the first trajectory is thus:

$$\text{Eq. 19} \qquad -\Delta E_{cycle} \approx 4\tau_1 \frac{dE}{dt}\bigg|_{m_x=1} + 2\tau_2 \frac{dE}{dt}\bigg|_{m_x=0} + 4\tau_3 \frac{dE}{dt}\bigg|_{m_z^{max}}$$

which reduces to:

$$\text{Eq. 20} \qquad \frac{\Delta E_{cycle}^i}{\mu_0 M_S^2} \approx \alpha \left( \frac{4h^2}{\sqrt{h_k}} + 2N_z^2 \sqrt{2h - h_k} + 2h/\sqrt{N_z h_k} \right)$$

The second and following i$^{th}$ trajectories have an incursion reduced by a factor $m_x^{max, i} < 1$ such that the dissipation energy of the i$^{th}$ trajectory is reduced by a factor $\left(m_x^{max, j}\right)^3$

In Eq. 20, the third term is by far the dominant contribution to the damping for near bifurcation trajectories, the case we are concerned with. The second and the third terms are comparable only if $h$ approaches $h_k$. The first term is negligible. Note that the above expression is valid only for $h_k/2 < h < h_k$. The third term would be to be discarded if the case $h > h_k$ was considered.

II.B. **Number of trajectories gone through before bifurcation**

The (non integer) number of trajectories NT gone through before bifurcation occurs is thus:

$$\text{Eq. 21} \qquad \sum_{i=1}^{NT} (m_x^{max\,j})^3 = \frac{\Delta E_{bif}}{\Delta E_{cycle}} \approx \frac{1}{\alpha} \times \frac{(h - h_k/2)\sqrt{N_z h_k}}{2h} \quad \text{for } h_k/2 < h < h_k$$

The physical meaning of the latter equation is transparent. The number of trajectories gone through before bifurcation occurs is the "excess field" ($h-h_k/2$) above the bifurcation criteria divided by some energy loss rate $2\alpha h/\sqrt{N_z h_k}$ occurring when the demagnetizing field is maximum during $\tau_3$, and taking into account that the total $(x)$ incursion reduces at each new trajectory gone through. The difficulty is to estimate the sum in Eq. 21. However since for all $i \in \{1..NT\}$ we have $\sqrt{1 - h^2/h_k^2} \approx \sqrt[3]{4} < m_x^{max} \leq 1$, the number of trajectories NT is not far from the right hand side of Eq. 21 as long as $2h-h_k << 1$.

It is worth noticing that $NT \times \alpha$ is almost a constant quantity. As expected, the smaller $\alpha$, the more trajectories the system goes through before being trapped.

The characteristic time $\tau_{bif}$ it takes for the system to bifurcate and round around a single attraction point is the number of trajectories divided by the slowest characteristic frequency ($\tau_2$), plus the delays:

$$\text{Eq. 22} \qquad \tau_{bif} \approx \frac{\sqrt{2}}{4\alpha\gamma_0 M_S} \frac{\sqrt{N_z h_k}\sqrt{2h - h_k}}{h} + \tau_1 + \tau_3$$

From Eq. 22, when very near the bifurcation criterion (i.e. when $2h - h_k << h$), the trapping process is immediate after the delays. The slight dissipation suffices to make the trajectory bifurcate and prevent the switching. The theoretical minimal switching field $h=h_k/2$ is thus out of reach as soon as $\alpha>0$.

II.C. Relaxation dominated precessional switching

Our first comment concerns the bifurcation singularity. Since it is the minimal cost in applied field for a switching event, it could seem interesting to try to switch the magnetization using a trajectory as close as possible to the bifurcation trajectory.
A important fact is that if $h_k$, $h > h_k/2$ and the damping $\alpha$ are such that the hard axis is passed by once, but may not be passed by a second time because of damping, the system falls in the nearest and thus most attractive half $m_x$ space. The reversal is achieved, even if the pulse is let on longer than $\tau_{bif}$.
Such a behavior was already obtained both experimentally [4] and with numerical simulations. Formerly reported as "relaxation dominated reversal", the present authors prefer to quote it as the *relaxation dominated precessional reversal*, to recall that its nature is very different from the Stoner-Wohlfarth astroïd (relaxation dominated) reversal.
The requirements on $h$, $h_k$ and $\alpha$ are given by Eq. 23 and displayed in Figure 4.

$$\text{Eq. 23} \qquad \alpha = \left(1 - \frac{h_k}{2h}\right)\sqrt{N_z h_k}$$

The previous equation is obtained considering [24] that the number of turns should be $NT = \frac{1}{2}$ for a full switch. With finite damping, the minimal switching field $h_{min}$ is:

$$\text{Eq. 24} \qquad h_{min} = \frac{h_k}{2} \times \left(\frac{1}{1 - \alpha/\sqrt{N_z h_k}}\right) \cong \frac{h_k}{2} + \frac{1}{2}\alpha\frac{\sqrt{h_k}}{\sqrt{N_z}}$$

The approximate expression holds in the limit of soft thin element, i.e. for $\alpha \ll \sqrt{N_z h_k}$. Note that the condition in the above equation is not stringent: practical MRAM cell exhibit $h_k$ typically less than 0.01, such that they can be considered as rather soft if their damping constant is much below 0.1, which is always the case in practice.

Another important conclusion is that there is an affine dependence of the minimal switching field with the damping parameter. Surprisingly, the proportionality constant is the half *square root* of the anisotropy field $H_k$.

## II.D. Discussion

The minimal field required for a relaxation dominated precessional switching event had been already calculated in [13] for a few discrete values of $\alpha$ and for $H_k$=13 kA/m. There (exact) numerical integration give an influence of $\alpha$ on the minimal cost $h_{min}$ which is compatible with Eq. 24.

The slight differences between exact numerical integration of LLG and our analytical study stem from our crude estimate of the term $\left(\sum_{i=1}^{NT}(m_x^{\max j})^3\right)/NT$ used to obtain Eq. 22.

For this reason, we calculated *numerically* the effect of $\alpha$ on the minimal switching field $h_{min}$, to further ameliorate our estimate of the prefactor of $\alpha$ in Eq. 23 and Eq. 24. As shown in Figure 4, putting an empirical prefactor of 0.59 instead of 0.5 in Eq. 24 gives a better overall agreement with the (exact) results produced using direct numerical integration of the LLG equation. The minimal switching field can be satisfactorily accounted for by a modified Eq. 24:

$$\text{Eq. 25} \qquad H_{min} = \frac{H_k}{2} + 0.59\frac{\alpha\sqrt{H_k}}{\sqrt{N_z M_S}}$$

which summarizes the effect of the damping parameter on the minimal field required to switch the magnetization in a precessional way.

## III. Conclusion

We have studied both analytically and numerically the precessional switching of an anisotropic, thin film nanostructure subjected to an in-plane applied field perpendicular to its easy axis. In the absence of damping, the exact magnetization trajectories could be derived. They were classified in anisotropy dominated trajectories, for which the switching is forbidden by energy considerations, and field dominated trajectories, for which switching is possible. Above that so-called bifurcation field, equal to half of the effective anisotropy field (Eq. 10), the characteristic switching frequency scales with the square root of the magnetization times the applied field minus half of the anisotropy field. In addition, this bifurcation trajectory has a frequency singularity, such that attempting to approach this limit makes the reversal intrinsically unpredictable because very sensitive to any noise or fluctuation source.

For larger applied fields, the system may switch several times back and forth. Several trajectories can be gone through before the system has dissipated enough energy to converge to one attracting equilibrium state. For some moderate fields, the system switches only once by a relaxation dominated precessional switching.

Finite damping allows designing precessional reversal strategies which succeed if the duration of the applied field is greater than a threshold. The corresponding minimal cost in applied field increases linearly with the damping parameter. The slope of this increase scales with the square root of the anisotropy field (Eq. 25). We believe that the simple and analytical charts derived in this paper may be useful guidelines for the design of magnetic random access memories switched by a precessional strategy.

Acknowlegment: The authors are grateful to Y. Suzuki who wrote the core of the LLG solving code used throughout this study.

**Figure captions**

Figure 1 : Exact magnetization trajectory of a non lossy macrospin film of initial magnetization along *(x)* when subjected to a transverse field *h=0.01* applied along *(y)*, with $N_z=1$. The uniaxial anisotropy along *(x)*. The anisotropy field is varied from $h_k$=*0.01, 0.016, 0.02* (twice the applied field: bifurcation trajectory), *0.03* and *0.1*. The two latter $h_k$ values correspond to anisotropy-dominated trajectories.
**A:** Trajectory in the *(zy)* plane
**B**: Trajectory in the *(xz)* plane
**C**: Vector **m** trajectory when initial magnetization is along *(x)*
Inset: Definition of the axes. The ellipse stands for the anisotropic macrospin.

Figure 2 : Magnetization trajectories of a thin film with uniaxial anisotropy $h_k$=*0.016* along *(x)* when a transverse field is apply along *(y)* with magnitude *h=0.01*. The initial magnetization $m_{y0}$ is varied between –0.2 and 0.5. The equilibrium position at infinite time would correspond to $m_{y0}$=*0.625*. Only one quadrant of the trajectory is displayed.

Figure 3 : Characteristic timescales of a precessional switching event.
**A**: Numerical integration of the $m_x$ component for $N_z$=*1*, $h_k$ = *0.0148*, $\mu_0 M_S$=*1.08* T, *h* = *1.0005* $h_k/2$ and $\alpha$=*0*. The initial delay $\tau_1$, the maximum speed time $\tau_3$ and the slow-down time $\tau_2$ are qualitatively superimposed on the time evolution.
**B**: Dependence of those three times for applied fields between $h_k/2$ and $h_k$=*0.01*, for $N_z$=*1*.

Figure 4 : Effect of the damping parameter $\alpha$ on the minimal switching field for relaxation dominated precessional switching. The corrected analytical estimate (Eq. 25, full lines) is compared to the results of full numerical integration of the LLG equation (cross symbols) for several values of the anisotropy, ranging from $h_k$=*0.0037* to *0.058*, for $N_z$=*1*.

Fig. 1A, T. Devolder et al.

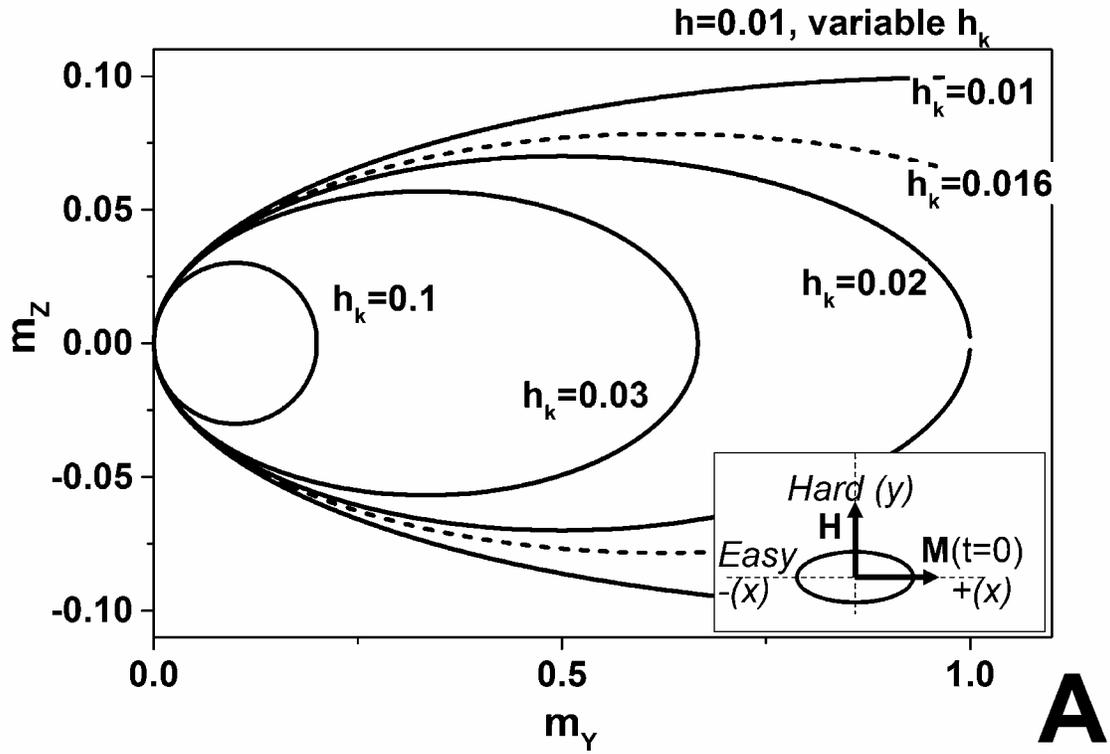

Fig. 1B, T. Devolder et al.

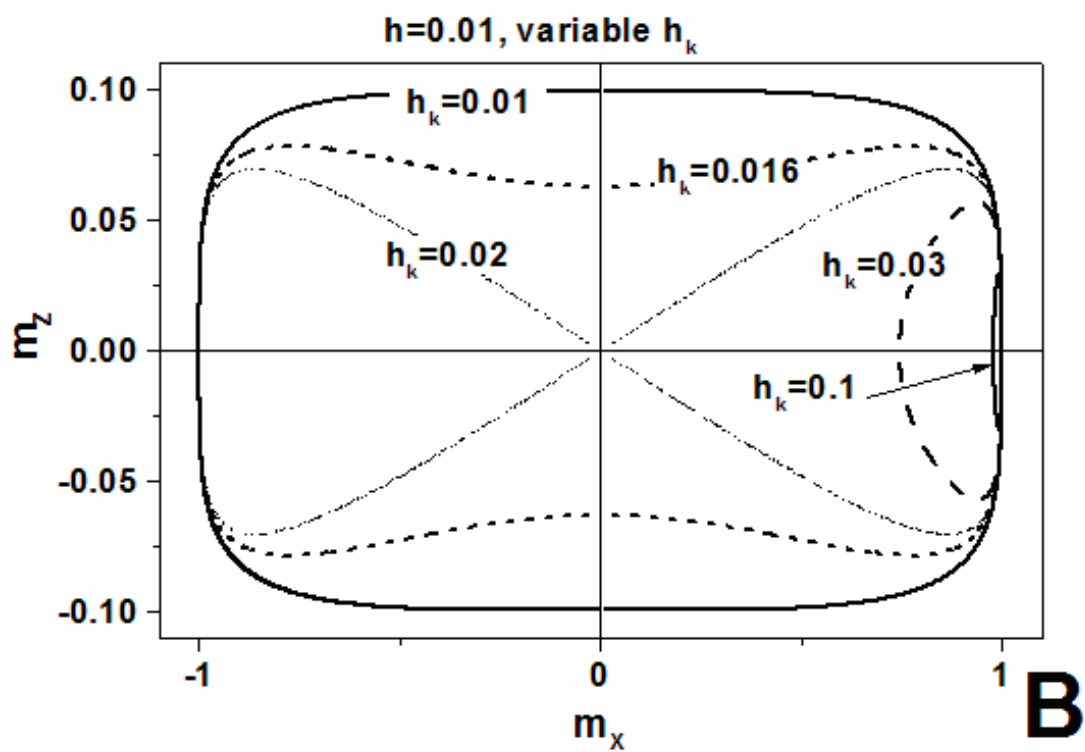

Fig. 1C, T. Devolder et al.

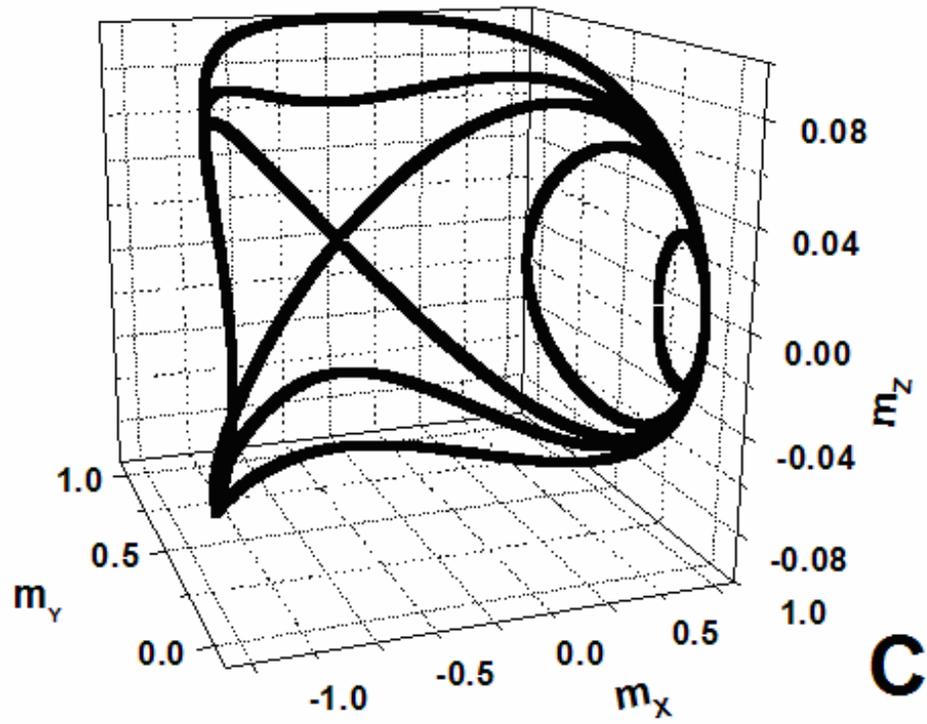

Fig. 2, T. Devolder et al.

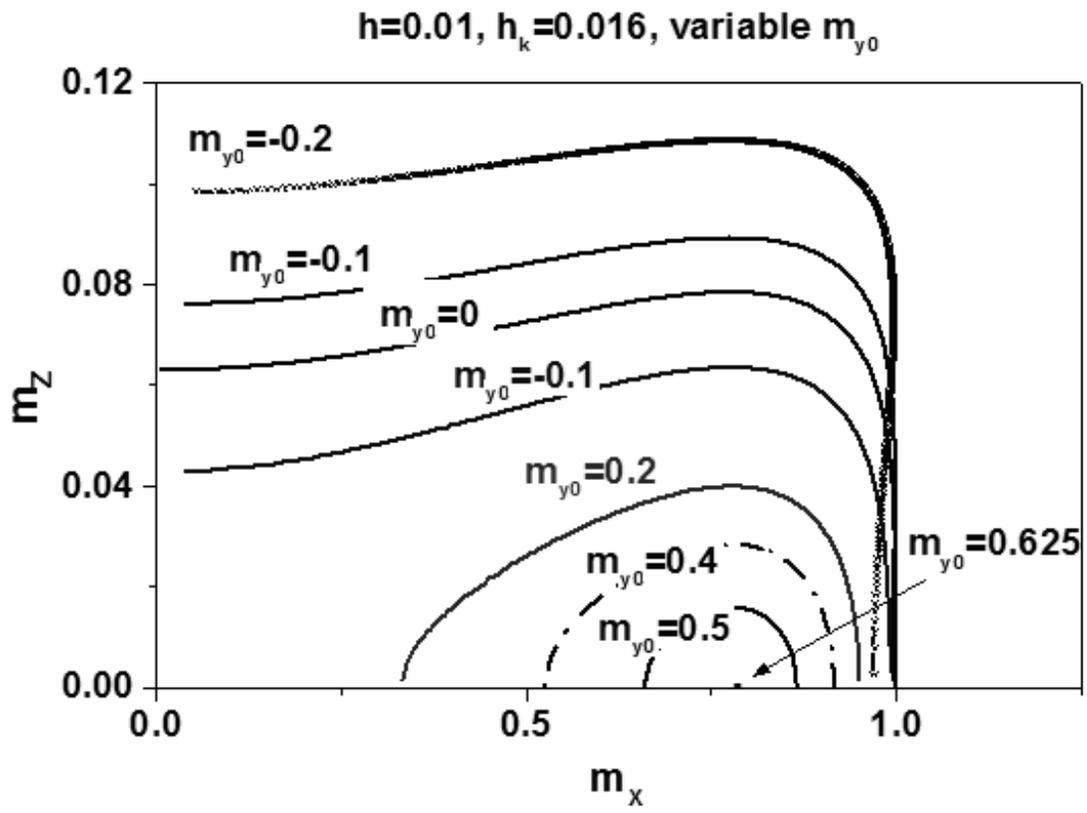

Fig. 3, T. Devolder et al.

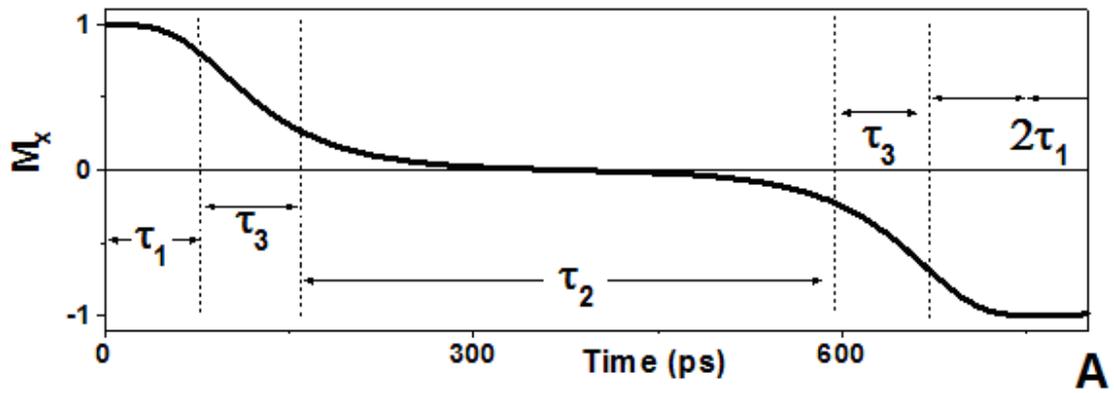

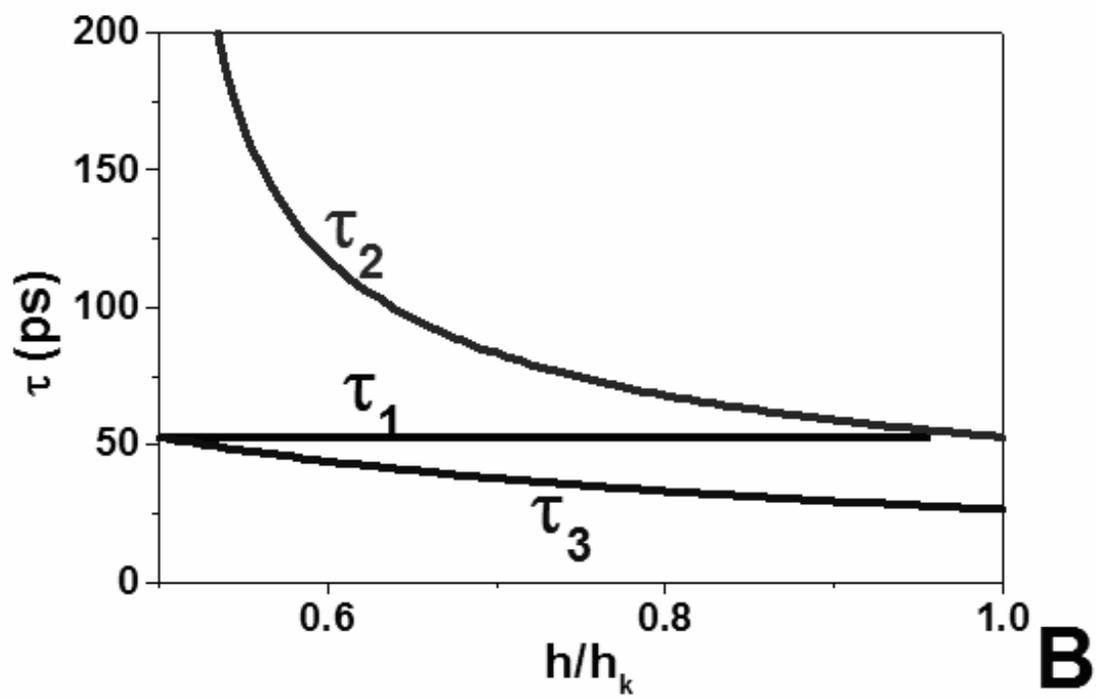

Fig. 4, T. Devolder et al.

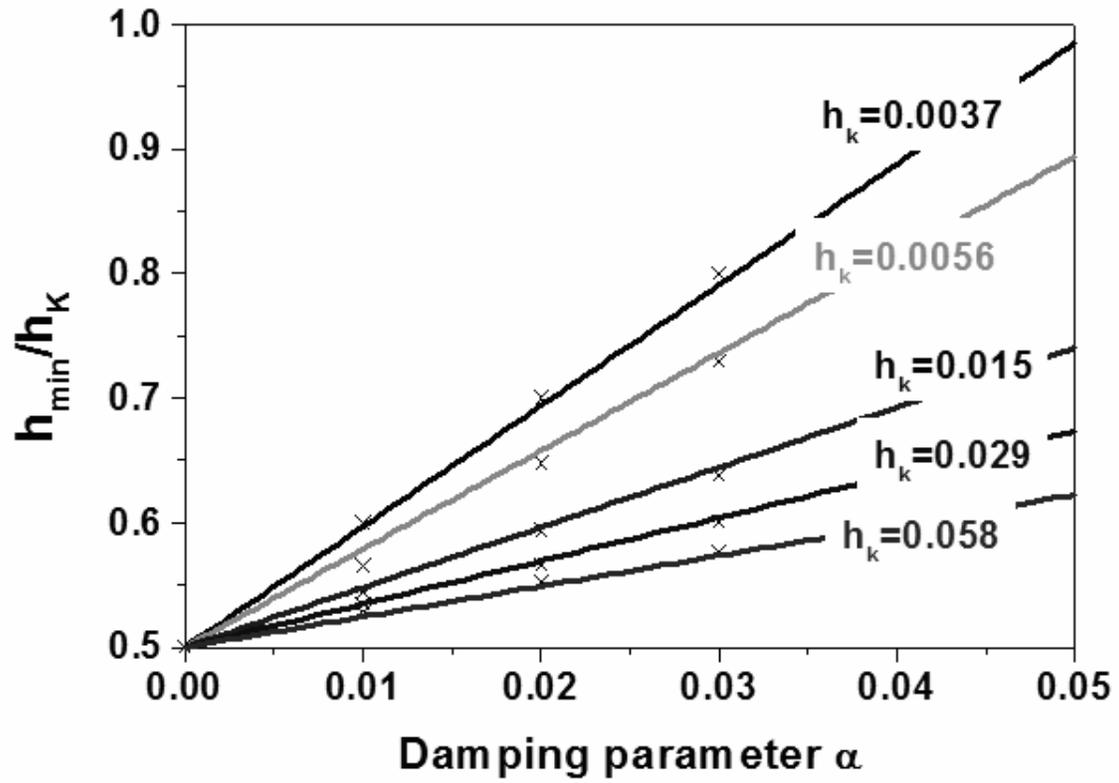